\documentclass{ptephy_v1}%%%%%% to generate preprint number with ptep logo

\usepackage[T1]{fontenc}
\usepackage{graphicx}                % figures
\usepackage{bm}                      % bold math
\usepackage{mathptmx}                % math+times fonts, after bm !
\usepackage{multirow}                % tables
\usepackage{dcolumn}                 % align table columns on decimal point
\usepackage{CJK}                     % chinese fonts
\usepackage{xcolor}                  % color
\def\bea{\begin{eqnarray}} \def\eea{\end{eqnarray}}
\def\be{\begin{equation}} \def\ee{\end{equation}}
\def\bal#1\eal{\begin{align}#1\end{align}}
\def\bse#1\ese{\begin{subequations}#1\end{subequations}}
\def\rra{\right\rangle} \def\lla{\left\langle}

\def\rv{\bm{r}}
\def\Jv{\bm{J}}
\def\zv{\bm{0}}
\def\eps{\varepsilon}
\def\la{\Lambda}
\def\kni{KNI} %\def\kni{$\bar{K}N$ interaction}

\def\fm3{$\,\text{fm}^{-3}$}
\def\mfm{$\,\text{MeV}\,\text{fm}^{-3}$}
\def\nspher{$^{16}_{\ K}$O, $^{40}_{\ K}$Ca, $^{208}_{\hspace{.7em}K}$Pb}
\def\ndef{$^{\,8}_{K}$Be, $^{20}_{\ K}$Ne, $^{36}_{\ K}$Ar}

\begin{document}

\begin{CJK}{UTF8}{gbsn}

\title{Deformed $\bm K^-$ nuclei in the Skyrme-Hartree-Fock approach}

\author{Yun Jin (金芸)}
\author[1]{Chao Feng Chen (陈超锋)}
\author[1,*]{Xian-Rong Zhou (周先荣)} \email{xrzhou@phy.ecnu.edu.cn}
\author[1]{Yi-Yuan Cheng (程奕源)}
\affil{
School of Physics and Materials Science, East China Normal University,
Shanghai 200241, P.~R.~China}

\author{H.-J. Schulze}
\affil{
INFN Sezione di Catania, Dipartimento di Fisica, Universit\'a di Catania,
Via Santa Sofia 64, 95123 Catania, Italy}

\begin{abstract}
The properties of kaonic nuclei are studied using a two-dimensional
Skyrme-Hartree-Fock model with a $KN$ Skyrme force.
We focus in particular on the instability of the solutions
for a too strong $KN$ interaction,
which determines a maximum value of the kaon binding in this approach.
We then analyze the change of the deformation properties of
several core-deformed nuclei caused by the added kaon,
and find a shrinking of the core
and in some cases a complete loss of deformation.
\end{abstract}

\subjectindex{D01}

\maketitle

\end{CJK}

%-------------------------------------------------------------------------------
\section{Introduction}

Kaonic nuclei are one of the important problems in the study of
strangeness physics \cite{rev,rev2,rev3,rev4}.
Experimental searches for $K^-$-nuclear bound states using
stopped kaon reactions with neutrons or protons were conducted early at
KEK and at DA$\Phi$NE \cite{kek,exp2,exp3}.
In recent years, they were microscopically investigated using meson beams
by new and upgraded experimental facilities at J-PARC,
where the broad $K^-pp$ bound-state structure is examined by current
experiments \cite{nagae,e15}.
Future experiments are foreseen at FAIR and HIAF \cite{fut}.

One expects that theoretical calculations of kaonic nuclei
to fit these experimental results may help us further explore the
$K^-N$ interaction (KNI).
Kaonic nuclei have been theoretically investigated by different models, i.e.,
relativistic mean field model
\cite{rmf,gal1,gal2,gazda1,gazda2,gazda3,gazda4,zhong1,zhong2,yang},
$G$-matrix model
\cite{aka1,aka2,aka3},
chiral condensate model
\cite{muto1,muto2},
%phenomenological density-dependent optical potential model \cite{friedman1},
self-consistent meson-baryon coupled-channel interaction model
\cite{weise1,weise2,weise3,weise4,hyodo,myo,mbcc1,mbcc2,mbcc3,mbcc4,mbcc5,
hrtan,hrtan2,ramos1,ramos2,ramos3,ramos4,seki},
Skyrme model
\cite{zhou},
etc.
Further constraints have been obtained by the study of
electromagnetically bound kaonic atoms
within the
phenomenological density-dependent optical potential model \cite{friedman1}.

In the relativistic mean field model,
several $K^-$ nuclei from $^{12}_{\ K}$C to
$^{208}_{\hspace{.7em}K}$Pb were studied
and the kaon binding energies of the $1s$ state were obtained
in the range of $49\sim76$ MeV \cite{rmf}.
Within the framework of antisymmetrized molecular dynamics, %(AMD),
kaonic nuclei were calculated by a $G$-matrix approach \cite{aka1},
and $^6_K$Be, $^8_K$Be, and $^9_K$B
were predicted as deeply bound kaonic nuclei.
Within the relativistic mean field theory combined with a chiral model,
the proton, neutron, and kaon density distributions of $^{15}_{8K}$O,
and its binding energy per kaon were calculated
at two values of the $K^-$ potential depth,
$V_K=-80$ MeV and $V_K=-120$ MeV \cite{muto2}.
The results indicated that the nuclear density in
the central region of multi-$K^-$ nuclei is saturated
for strangeness $|S|\geqslant8$.
In the self-consistent meson-baryon coupled-channel interaction model,
the $K^-pp$ structure was found as a shallowly bound system,
and the binding energy of the $K^-pp$ was obtained in the range
of $14\sim50$ MeV \cite{weise1}.
The phenomenological density-dependent optical potential model predicts the
deepest values of $150\sim200$ MeV for the $K^-$ potential depth
\cite{friedman1,friedman2,friedman3,friedman4,friedman5}.

However, there is still much controversy on the $K^-$-nucleus
bound state and the depth of the kaon nuclear optical potential.
Theoretical coupled-channel calculations employing only (chiral)
two-body forces yield widely varying results
for the real and imaginary part of the optical potential \cite{hrtan,hrtan2}.
The recent $K^-$-nuclear experiments investigate the possible existence
of a deeply bound $K^-pp$ state \cite{nagae,e15}
in order to obtain experimental constraints on this problem,
but currently the issue is still unsolved.

In 2005, the first evidence of a kaon bound state $K^-pp$ was observed
through its decay into a proton and $\Lambda$ at DA$\Phi$NE
by the FINUDA Collaboration \cite{exp2}.
They deduced a binding energy of
$115^{+6}_{-5}$(stat.)$^{+3}_{-4}$(sys.)\,MeV,
as well as a decay width of $67^{+14}_{-11}$(stat.)$^{+2}_{-3}$(sys.)\,MeV.
In 2010, a deeply-bound $K^-pp$ state was indicated by the analysis of the
DISTO experiment, which studied the intermediate $K^-pp$ state in the
$p+p \rightarrow K^-pp + K^+ \rightarrow p + \Lambda + K^+$
reaction \cite{Yamazaki}.
But later it was questioned by the HADES Collaboration's re-analysis
of the $p+p$ reaction data \cite{Epple}.
Recent experiments at J-PARC by the E15 and E27 collaborations
presented latest results.
E27 claimed that they had observed a deeply bound state $K^-pp$,
which was produced in the
%$d(\pi^{+}, K^{+})$ reaction at 1.69 GeV/c,
$\pi^+ + n+p \rightarrow K^-pp + K^+ \rightarrow \Sigma^0 + p + K^+$ reaction.
And the E15 collaboration observed a bound state $K^-pp$ with a binding energy
of $47\pm3(\text{stat.})^{+3}_{-6}(\text{sys.})\,$MeV \cite{nagae}.

In Ref.~\cite{zhou} we introduced the
Skyrme-Hartree-Fock (SHF) approach for the study of kaonic nuclei,
namely we combined a standard $NN$ Skyrme force with an effective \kni\
obtained within the chiral model of \cite{muto1}.
The purpose of our present paper is to continue this study by replacing
the chiral \kni\ with a simpler but more general
phenomenological $KN$ Skyrme force,
and in particular to investigate deformed nuclei within a 2D Skyrme calculation.
The motivation is that a rather strong \kni\ might cause
profound changes of the nuclear core structure in some cases,
which can be studied and understood in detail in our approach.
In particular,
we will consider the $K^-$ nuclei with spherical core
$^{16}_{\ K}$O ($^{16}$O+$K^-$),
$^{40}_{\ K}$Ca ($^{40}$Ca+$K^-$),
$^{208}_{\hspace{.7em}K}$Pb ($^{208}$Pb+$K^-$),
and the deformed-core nuclei
$^{\,8}_{K}$Be ($^{8}$Be+$K^-$),
$^{20}_{\ K}$Ne ($^{20}$Ne+$K^-$),
$^{36}_{\ K}$Ar ($^{36}$Ar+$K^-$).

The paper is organized as follows.
In Sec.~II, the self-consistent SHF approach with a simple Skyrme force for the
\kni\ is presented.
Sec.~III shows the obtained results
and discussions of the neutron, proton, and kaon mean fields and density
distributions for the spherical $K^-$ nuclei \nspher.
In Sec.~IV the deformation and potential energy surfaces of \ndef\
are analyzed and compared with their core nuclei.
Finally, we make a summary in Sec.~IV.

%-------------------------------------------------------------------------------
\section{Formalism}

Our calculation is performed in the SHF approach with a
density-dependent Skyrme force for the \kni.
In this approach, the total energy of a nucleus is written as
usual \cite{vaut,vautdef,rayet,rayet2,shfrev1,shfrev2,shfrev3,hypsky} as
\be
 E = \int d^3\rv\, \eps(\rv) \ ,\quad
 \eps = \eps_{NN} + \eps_{KN} + \eps_C \:,
\ee
where $\eps_{NN}$ is the nucleon-nucleon part of the energy-density functional,
$\eps_{KN}$ is the kaon-nucleon part,
and $\eps_C$ is the electromagnetic part due to the Coulomb interaction
of protons and kaons.
These energy-density functionals
depend in general on the one-body density $\rho_q$,
kinetic density $\tau_q$, and spin-orbit current $\Jv_q$
(only for nucleons),
\be
  \big[ \rho_q,\; \tau_q,\; \Jv_q \big] =
  \sum_{i=1}^{N_q} {n_q^i} \Big[
  |\phi_q^i|^2 ,\;
  |\bm\nabla\phi_q^i|^2 ,\;
  {\phi_q^i}^* (\bm\nabla \phi_q^i \times \bm\sigma)/i
 \Big] \:,
\ee
where $\phi^i_q$ ($i=1,N_q$) are the
self-consistently calculated single-particle (s.p.) wave functions
of the $N_q$ occupied states for the different particles
$q=n,p,K^-$ in a nucleus.

The minimization of the total energy implies the SHF Schr\"odinger equation
for each single-particle state,
%$i$ and $q=n,p,K$,
\be
 \Big[ -\bm\nabla\cdot\frac{1}{2m^*_q(\rv)}\bm\nabla
 + V_q(\rv) - \bm W_q(\rv)\cdot(\bm\nabla\times\bm\sigma)
 \Big] \phi_q^i(\rv)
 = e_q^i\phi_q^i(\rv)
\label{e:seq}
\ee
with the mean fields
(including the Coulomb interaction)
\bal
 V_K &= \frac{\partial\eps_{KN}}{\partial\rho_K} - V_C \:,
\\
 V_q &= V_q^\text{SHF} \!+ V_q^{(K)} \ ,\
 V_q^{(K)}=\frac{\partial\eps_{KN}}{\partial\rho_q}
 \ ,\ (q=n,p) \:.
\eal
The spin-orbit mean field $\bm W_{n,p}$ is the one of the $NN$ Skyrme force
used here,
and we put $\bm W_K=0$ in this study.

For $\eps_{NN}$ we use a standard nucleonic Skyrme energy density
functional \cite{shfrev1,shfrev2,shfrev3}
(in this paper the SLy4 parametrization \cite{sly,sly2,sly3}),
depending on the densities $\rho_{n,p}$, $\tau_{n,p}$, $\Jv_{n,p}$,
and for the kaonic contribution we make the simple ansatz
\be
 \eps_{KN} = -a_0 \rho_K \big[(1+x_0)\rho_p+(1-x_0)\rho_n \big] \:,
\label{e:eps}
\ee
in view of the fact that currently not even the magnitude of the
\kni\ is well known.
With this definition in nearly symmetric systems,
$\rho_p\approx\rho_n$,
the results depend primarily on $a_0$ and much less on $x_0$.
For a rather strong asymmetric KNI,
the $n$ and $p$ density distributions will be affected differently,
and the dependence on $x_0$ will become apparent.
These features will be seen later.

We take into account pairing forces (between nucleons only)
within BCS approximation,
employing a density-dependent $\delta$ force \cite{nta},
\be
 V_q(\rv_1,\rv_2) = -V'_q
 \Big[ 1 - \frac{\rho_N((\rv_1+\rv_2)/2)}{0.16\,\text{fm}^{-3}} \Big]
 \delta(\rv_1-\rv_2) \:
\ee
with $V'_n=V'_p=410$\mfm\ as pairing strength in light nuclei
\cite{hsa1,hsa2,hsa3}
and $V'_p=1146$\mfm, $V'_n=999$\mfm\ for medium-mass and heavy nuclei.
A smooth energy cutoff is included in the BCS calculation \cite{mbe}.

In the following we will study the dependence of the main observables,
such as kaon mean field $V_K$,
kaon removal energy
\be
 B_K \equiv E(^AZ) - E(^A_KZ)
\ee
(in this notation
$A$ is the nucleon number and
$Z$ is the proton number, not the charge number)
and also deformation properties of some nuclei,
on the KNI strength parameter $a_0$.

In microscopic approaches for the KNI \cite{muto1,hrtan,hrtan2},
the $K^-p$ interaction
is usually much more important (attractive) than the $K^-n$ interaction,
and we take account of this fact by comparing in the following the two extreme
choices $x_0=0$ and $x_0=1$,
respectively modeling a ($p$,$n$)-symmetric \kni\ and the case of
neglecting the $K^-n$ interaction completely,
while doubling the $K^-p$ interaction.
We will study whether the two choices lead to significantly different predictions
for some observables.

The choice of the KNI functional Eq.~(\ref{e:eps})
amounts to a simple linear density dependence of the kaon mean field,
\bal
 V_K &= -a_0 \big[(1+x_0)\rho_p+(1-x_0)\rho_n \big] - V_C \:,
\\
 V_{\scriptstyle p \atop \scriptstyle n}^{(K)} &=
 -a_0(1 \pm x_0)\rho_K \:.
\eal
In the future, once enough reliable data become available,
the functional can of course be extended by
adding nonlinear density dependence, surface terms, etc.,
as in the case of the SHF approach for
$\la$ hypernuclei \cite{rayet,hypsky}.
At the moment, it is clearly premature
to determine all the \kni\ parameters of these terms.

A similar remark concerns the imaginary part of the \kni\,
due to the decay channels
$KN \rightarrow \pi Y$,  %\;Y=\la,\Sigma$
$KNN \rightarrow YN$
($Y=\la,\Sigma$) \cite{seki,hrtan,hrtan2}.
%$KN \rightarrow \la(1405) \rightarrow \pi\Sigma$
While we study here the effect of the KNI strength parameters $a_0$ and $x_0$
of the real part $\text{Re}\, V_K$
on the instability and properties of kaonic nuclei,
we currently neglect the imaginary part in attendance of reliable data.
It has been found that the effect of a moderate
$\text{Im}\,V_K\lesssim 20$ MeV
(neglecting the kaon multinucleon absorption)
on the real part is nearly negligible,
whereas too large widths might make kaon bound states
unobservable \cite{hrtan,hrtan2}.
We consider this feature an open problem
that can only be solved by future confrontation with accurate data.
We give a brief estimate of the qualitative effect in our formalism
in the next section, though.

Assuming axial symmetry of the mean field, the deformed SHF Schr\"odinger
equation is solved in cylindrical coordinates $(r,z)$ within the
axially-deformed harmonic-oscillator basis
\cite{vautdef,shfrev1,shfrev2,shfrev3}.
This allows to model axially-deformed nuclei,
which will be discussed in the following.

%-------------------------------------------------------------------------------
\section{Results}

\begin{table}[t]%...............................................................
\caption{
The central kaon potential $V_K \equiv -V_K(\rv=\zv)$
and the kaon separation energy $B_K$ for the (spherical) ground states of
several nuclei in the case of $a_0=300$, 700 \mfm\ and $x_0=0,1$.
The maximum values of $a_0$ for stable calculations
and the corresponding $V_K$ and $B_K$ are also listed.
Energies are given in MeV and $a_0$ in \mfm.
}
\label{table}
\centering
%\begin{ruledtabular}
\begin{tabular}{ccccc|cc|cc}
\hline
&       &                  &                      &
& \multicolumn{2}{c|}{$a_0=300$} & \multicolumn{2}{c}{$a_0=700$} \\
%& $x_0$ & $a_0^\text{max}$ & $B_{\!K}^\text{max}$
& $x_0$ & $a_0^\text{max}$ & $V_{\!K}^\text{max}$ & $B_{\!K}^\text{max}$
& $V_K$  &  $B_K$                  &  $V_K$  &  $B_K$                 \\
\hline
\multirow{2}{*}
{$^{\,8}_{K}$Be}     & 0 & 740 & 291.9 &  77.9 & 61.6 &  9.2 & 257.1 &  68.7 \\
                     & 1 & 330 & 169.4 &  14.4 & 88.5 & 10.3 &       &       \\
\multirow{2}{*}
{$^{16}_{\ K}$O}     & 0 & 845 & 345.6 & 122.6 & 58.1 & 21.1 & 217.0 &  87.3 \\
                     & 1 & 359 & 150.1 &  31.6 & 73.0 & 22.1 &       &       \\
\multirow{2}{*}
{$^{20}_{\ K}$Ne}    & 0 & 807 & 289.6 & 116.8 & 57.4 & 25.0 & 204.9 &  92.3 \\
                     & 1 & 351 & 122.7 &  33.8 & 69.3 & 25.8 &       &       \\
\multirow{2}{*}
{$^{36}_{\ K}$Ar}    & 0 & 837 & 343.8 & 144.4 & 74.9 & 36.9 & 225.1 & 111.1 \\
                     & 1 & 376 & 170.6 &  50.9 & 86.5 & 37.2 &       &       \\
\multirow{2}{*}
{$^{40}_{\ K}$Ca}    & 0 & 867 & 352.0 & 146.1 & 71.5 & 37.4 & 208.9 & 107.5 \\
                     & 1 & 403 & 162.1 &  55.1 & 79.0 & 37.6 &       &       \\
\multirow{2}{*}
{$^{208}_{\ \ K}$Pb} & 0 & 902 & 362.2 & 165.4 & 75.2 & 62.2 & 160.8 & 126.8 \\
                     & 1 & 559 & 173.7 &  87.7 & 77.3 & 53.7 &       &       \\
\hline
\end{tabular}
%\end{ruledtabular}
\end{table}%....................................................................

\begin{figure}[t]%..............................................................
\vspace{-10mm}\hspace{8mm}
\centerline{\includegraphics[scale=0.35]{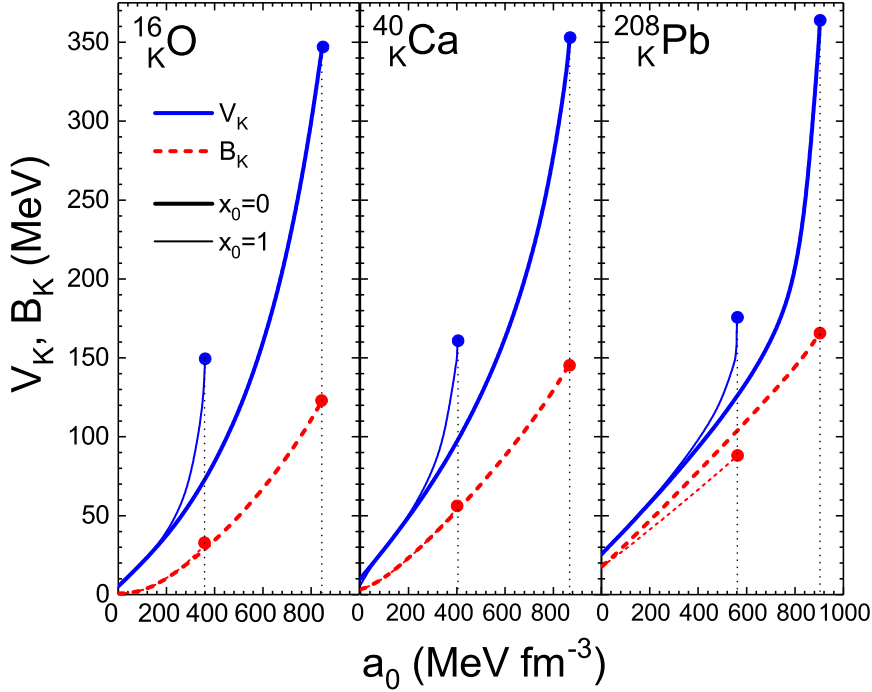}}
\vspace{-10mm}
\caption{
The central kaon potential $-V_K(\rv=\zv)$ (solid blue curves)
and the kaon removal energy $B_K$ (dashed red curves)
of \nspher,
calculated with the SLy4 $NN$ force and the $KN$ Skyrme force
for different interaction parameters $a_0$
and the two choices $x_0=0$~(thick curves), 1~(thin curves).
The markers indicate the onset of instability of the SHF solutions.
}
\label{f:va}
%\mycom{x0,x1 in black}
\end{figure}%...................................................................

\begin{figure}[t]%..............................................................
\vspace{-17mm}
\centerline{\includegraphics[scale=0.85]{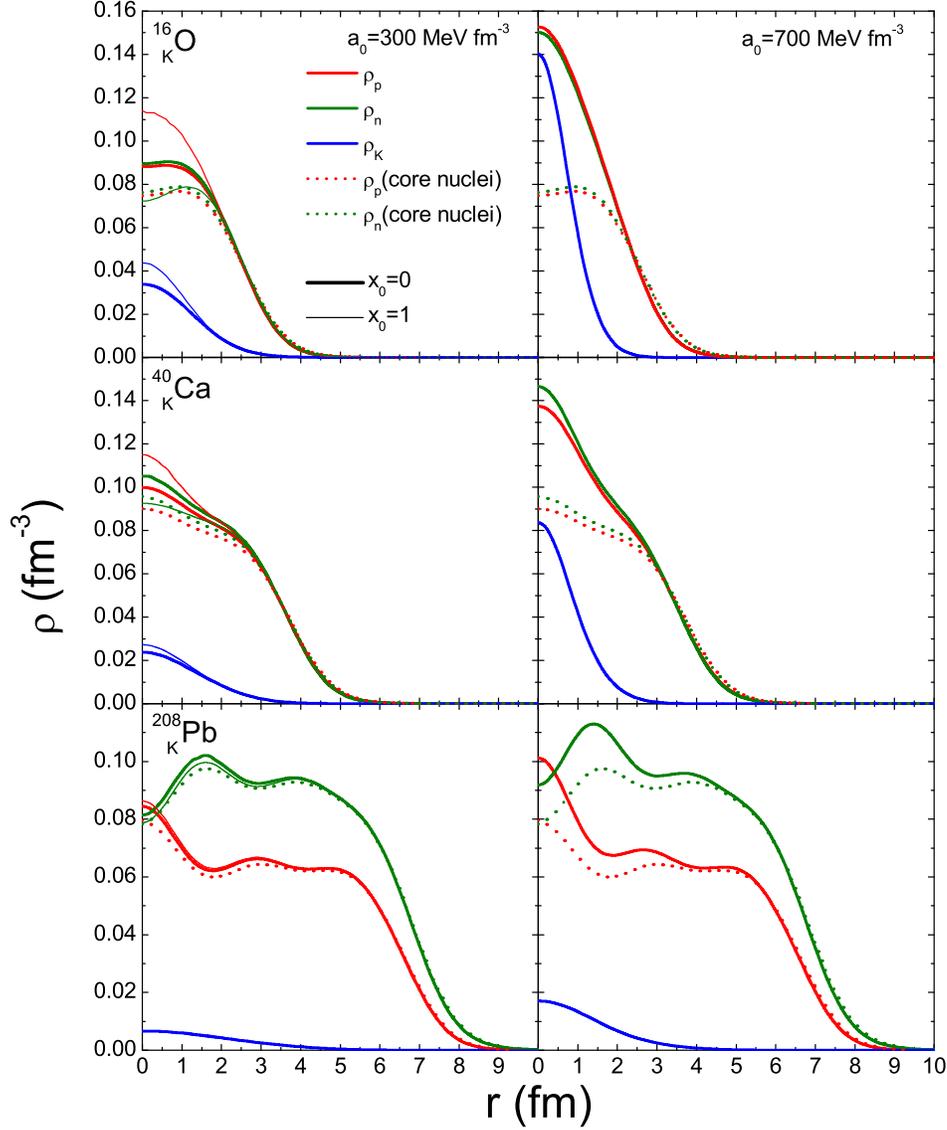}}
\vspace{-20mm}
\caption{
Proton, neutron, and kaon number density distributions of \nspher\
(solid curves)
and their corresponding core nuclei
(dotted curves)
in the case of $a_0=300$ (left panels), 700 (right panels) \mfm\
and for $x_0=0$~(thick curves), 1~(thin curves in left panels).
}
\label{f:rho}
%\mycom{K in blue}\\
\end{figure}%...................................................................

\begin{figure}[t]%..............................................................
\vspace{-3mm}
\centerline{\includegraphics[scale=0.62]{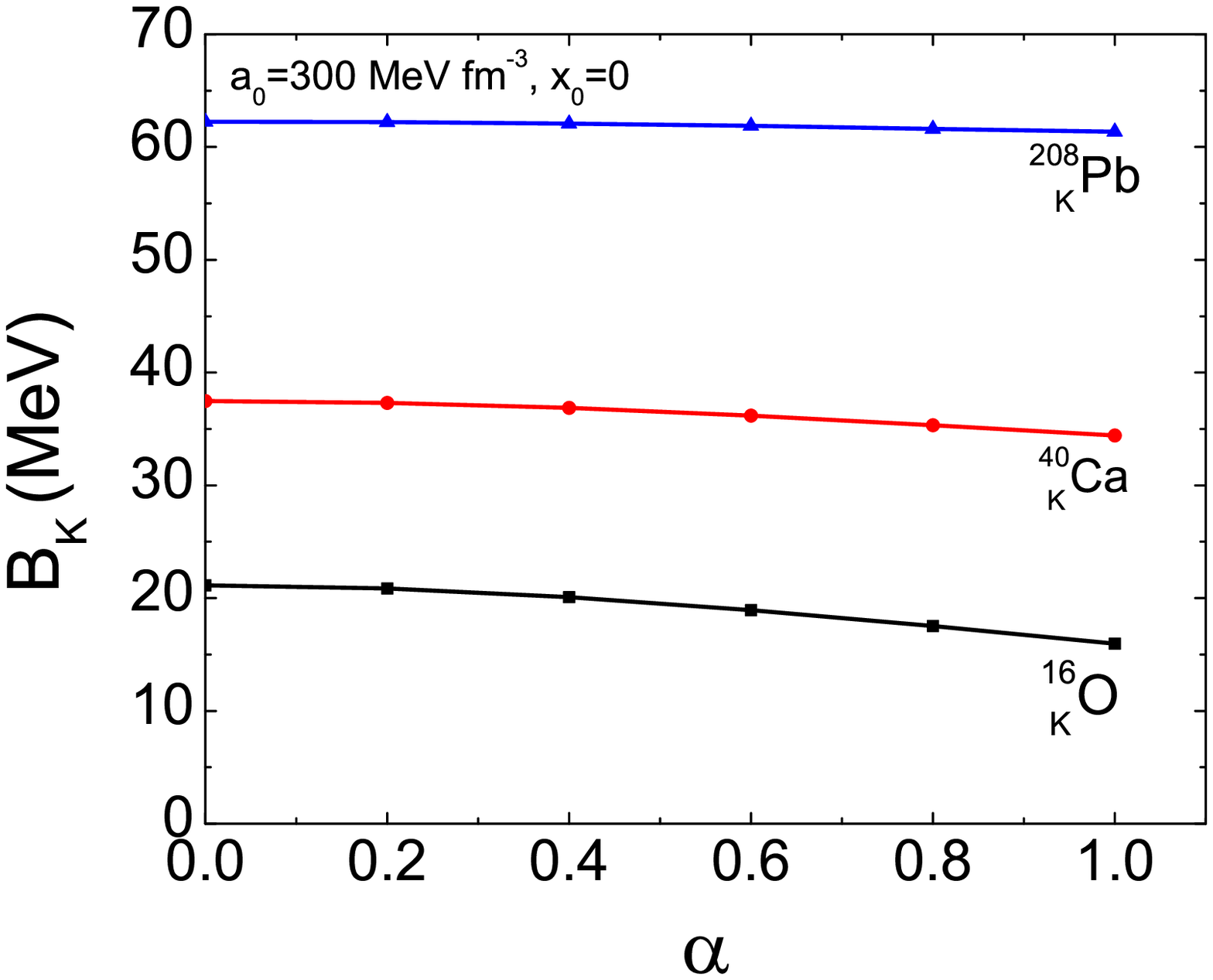}}
\vspace{-4mm}
\caption{
Kaon removal energy as a function of the absorption parameter $\alpha$,
Eq.~(\ref{e:a}),
for $a_0=300$\mfm, $x_0=0$,
and the nuclei \nspher.
}
\label{f:img}
\end{figure}%...................................................................

\begin{figure}[t]%..............................................................
\vspace{-1mm}
\centerline{\includegraphics[scale=0.68]{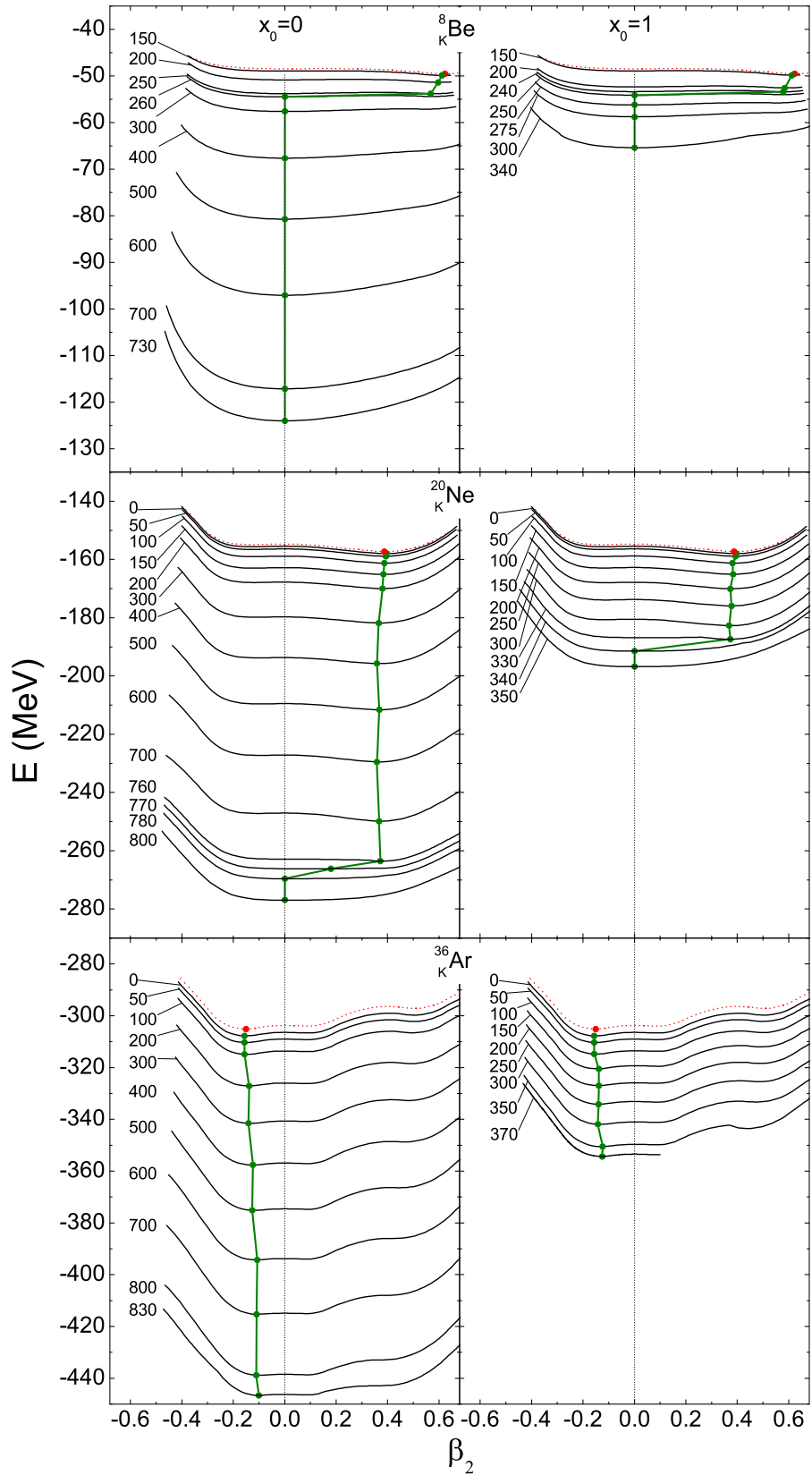}}
\vspace{-2mm}
\caption{
Potential energy surfaces of \ndef\
for different KNI strengths
$a_0$
(in \mfm, numbers near the curves)
%=0,100,\ldots,800$ \mfm\
and for
$x_0=0$~(left panels), 1~(right panels).
The dotted red curves are those of the core nuclei.
The minima are indicated by markers
and connected by a green line.
}
\label{f:def}
\end{figure}%...................................................................

\begin{figure}[t]%..............................................................
\vspace{-2mm}
\centerline{\includegraphics[scale=0.92]{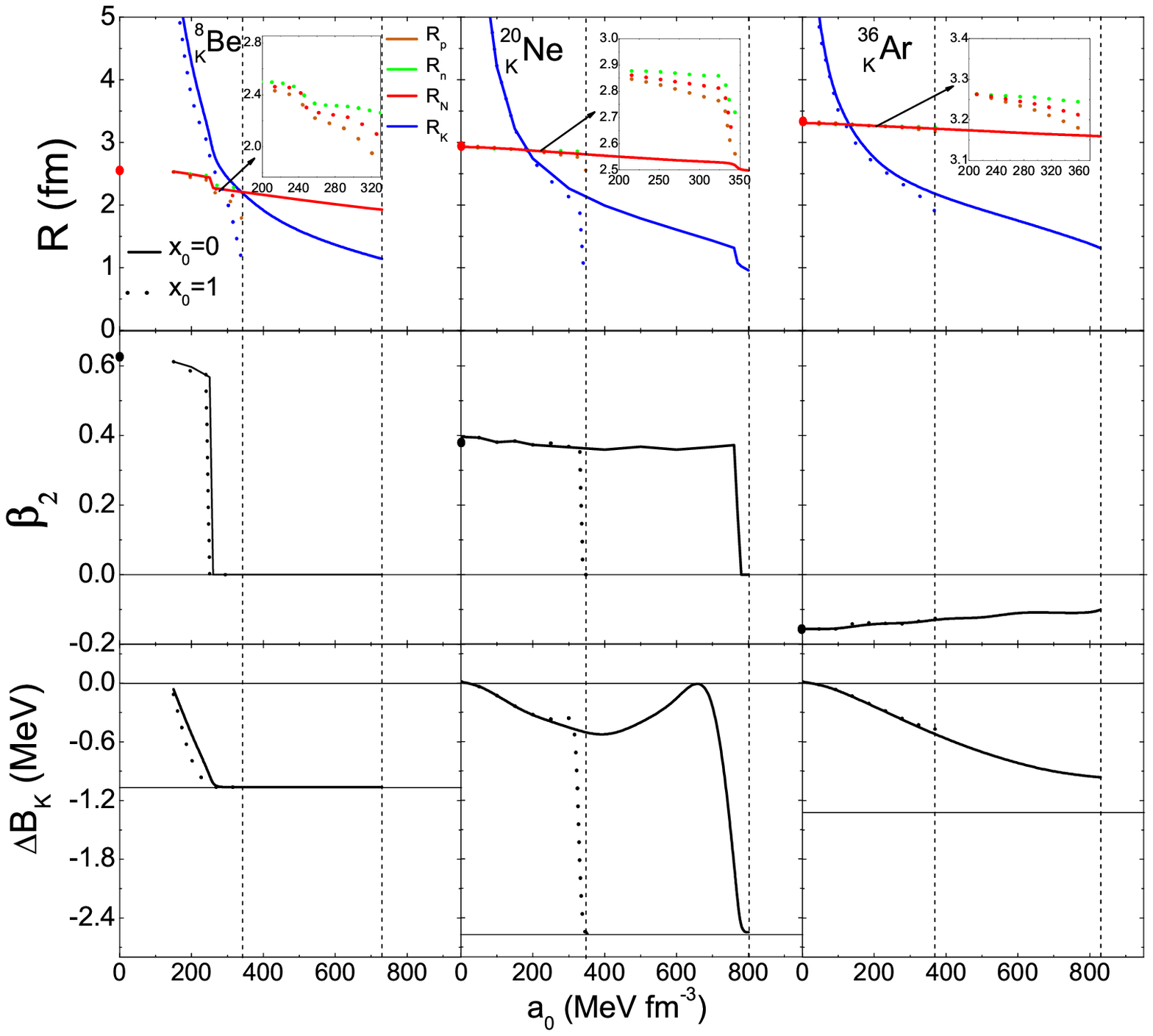}}
\vspace{-8mm}
\caption{
Rms radii $R_p$, $R_n$, $R_N$, $R_K$
(upper panels),
deformation parameter
%$\beta_2^{(p)}$, $\beta_2^{(n)}$,
$\beta_2$
(central panels),
and change of kaon removal energy due to deformation,
Eq.~(\ref{e:db}), (lower panels)
at the minimum of the BES
vs.~$a_0$ for \ndef\ and $x_0=0$, 1.
}
\label{f:ba}
\end{figure}%...................................................................

We now study the properties of kaonic nuclei with spherical cores, \nspher\
and deformed cores, \ndef.
We first consider $a_0$ as a free parameter
(for both choices $x_0=0,1$)
and study the strength of the binding as a function of its value.
Then we illustrate the change of nuclear structure in more detail
for two reasonable choices of $a_0$.

%-------------------------------------------------------------------------------
\subsection{Kaonic nuclei with spherical core \nspher}

Figure~\ref{f:va} shows the
central kaon potential $-V_K(r=0,z=0)$
and the kaon removal energy $B_K$
as a function of the \kni\ parameter $a_0$
for the three spherical kaonic nuclei \nspher.
One observes that both quantities increase
first linearly and then more rapidly with the \kni\ strength,
until meeting instability points at
$a_0=845$, 867, 902 \mfm\
($B_K=123$, 146, 165 MeV)
for $x_0=0$ and the three nuclei, respectively.
For the asymmetric KNI, $x_0=1$,
the instability occurs much earlier at
$a_0=359$, 403, 559 \mfm\
($B_K=32$, 55, 88 MeV).
We had encountered this phenomenon already in Ref.~\cite{zhou}:
For a too strong attractive \kni,
the central densities of nucleons and kaon may increase without limit,
and at a certain point the nuclear core is not stable any more against collapse.
This feature has been observed in Ref.~\cite{zhou}
for different nucleonic Skyrme forces,
which all yielded similar collapse points.
It might be cured by introducing suitable repulsive $NN$ and $KN$ forces
active at high density,
and the maximum of $a_0$ might then be enlarged.
But until that can be reliably done,
we consider the above limits on $a_0$ as the possible reasonable range
of our theoretical investigation.
%of the solution of stable kaonic nuclei.

The difference between the $x_0=0$ and $x_0=1$ interactions is large,
because in the latter case the proton component of the nucleonic core
is much more distorted for a given $a_0$ than in the former case,
and therefore instability sets in earlier.
Consequently maximum potential $V_K^\text{max}$
and kaon removal energy $B_K^\text{max}$
are much smaller than those for $x_0=0$.
Note that in particular for the light nuclei,
the possible value of the removal energy $B_K$ is severely limited in this case.
However, for a given value of $a_0$,
$B_K$ is nearly independent of the asymmetry parameter $x_0$.

In the following we fix the KNI parameter to two typical
`weak' and `strong' values
$a_0=300$ or 700 \mfm\
in order to investigate in more detail the changes of nuclear structure.
The corresponding values of $V_K$ and $B_K$
for the different nuclei
(and also for \ndef\ in spherical approximation)
are listed in Table~\ref{table}.
For example, in $^{40}_{\ K}$Ca the two choices correspond to
$V_K\approx 70$ and 200 MeV,
%which reflects more or less the range
compared to a range $V_K\approx\;$(30--110) MeV
currently obtained with more microscopic chiral forces
\cite{hrtan}. %for example.

For these parameter choices,
Fig.~\ref{f:rho} shows the density distributions in the three nuclei.
Obviously the effect of the inserted kaon is larger in light nuclei,
where the central proton and neutron distributions are substantially enhanced
due to its presence.
One can also see clearly the difference between neutron and proton core
distortions for the $x_0=1$ case, mentioned before.
We can therefore anticipate a substantial reduction of the nuclear rms
radius $R_N$, i.e.,
a shrinking of the nucleus.
This will be studied in more detail in the following.

Before that,
we return to the problem of the imaginary part of the kaon optical potential,
which we model by solving the SHF Schr\"odinger equation (\ref{e:seq})
incorporating a complex kaon potential,
\be
 V_K(\rv) = V_R(\rv) + iV_I(\rv) \:.
\ee
For simplicity we assume here a proportionality between both components,
\be
 V_I=-\alpha V_R \:,
\label{e:a}
\ee
in order to study the importance of the effect qualitatively.
The imaginary part modifies the kaon wave function, single-particle energy,
density distribution,
and therefore the kaon removal energy.
In Fig.~\ref{f:img} we display the
(real part of the)
kaon removal energy as a function of the
absorption parameter $\alpha$
for $a_0=300$\mfm, $x_0=0$, and the three nuclei \nspher.
One notes that the change of $B_K$ is small,
even up to a fairly large value $\alpha=1$.
Furthermore, a given value of $B_K$ could always be restored
by slighty adjusting the value of $a_0$ in this model.
This demonstrates that the imaginary part of the kaon mean field
does not play an important role in the SHF model,
at least regarding its effect on the real part and the kaon removal energy.
Treatment of real and imaginary part can be fairly well separated.
Of course more experimental information is required for a
final parameter fitting
beyond the simple proportionality assumption made here.

%-------------------------------------------------------------------------------
\subsection{Kaonic nuclei with deformed core \ndef}

The strong contraction of the nuclear core observed in Fig.~\ref{f:rho}
motivates the extension of our model to deformed nuclei.
In the case of $\la$ hypernuclei
(with a substantially smaller $\la N$ interaction strength
$a_0^{(\la N)}\approx 300\;$\mfm\ \cite{hypsky}
compared to the $KN$ one),
the modification of the nuclear core by the inserted $\la$
is a well known theoretical phenomenon \cite{tanida,shfdef,hagino,cdef},
namely, both the nuclear core radius
\be
 R_N \equiv \sqrt{ \lla r^2 + z^2 \rra }
 = \sqrt{ \frac{N}{A} \langle R_n^2 \rangle + \frac{Z}{A} \lla R_p^2 \rra}
\label{e:rn}
\ee
and the nuclear quadrupole deformation
\be
 \beta_2 \equiv \sqrt{\pi\over5}
 {\langle 2z^2-r^2 \rangle \over \langle r^2 + z^2 \rangle} \:
\label{e:beta2}
\ee
might be strongly affected by the added hyperon:
shrinking and reduction of core deformation, respectively.
We therefore carry out an equivalent analysis for kaonic nuclei now.
At variance with $\la$ hypernuclei, however, the kaon
decays on the same timescale as the eventual rearrangement
of the nuclear structure,
and this effect is neglected
together with the imaginary part of the KNI
in our present theoretical approach.
Only a future dynamical simulation of the nuclear rearrangement could
provide a more realistic picture.

Fig.~\ref{f:def}
shows the modification of
the ground-state potential energy surfaces (PESs) of \ndef\
with increasing value of the parameter $a_0$.
(The collapse points
are $a_0=$740, 807, 837 \mfm\ for $x_0=0$
and $a_0=$330, 351, 376 \mfm\ for $x_0=1$, respectively).
In general the PESs
become then increasingly flatter around the local minimum
compared to the corresponding core nuclei.
The deformation of the light nuclei $^{\,8}_K$Be and $^{20}_{\ K}$Ne
might even completely vanish due to the added kaon
for a sufficiently strong \kni,
$a_0>253 (243)$ \mfm
and $a_0>768 (333)$ \mfm, respectively,
whereas for the heavier extended $^{36}_{\ K}$Ar nucleus
the presence of a single kaon
concentrated in the center is not enough to eliminate the deformation
up to the highest possible values of $a_0$.
As a word of caution we remark that in general the predicted deformation
properties of the core nuclei
depend on the $NN$ Skyrme force that is employed \cite{cdef};
however, for the strongly deformed $^8$Be most Skyrme forces agree on a
$\beta_2\approx0.63$ \cite{shfdef}.
A further comment regards the mean-field approximation employed here,
which might be inadequate in particular for weak PES minima,
due to the neglect of configuration mixing.
A beyond-mean-field treatment \cite{cui1,cui2,cui3,mei}
might be required for a more realistic modelling.

Fig.~\ref{f:ba}
summarizes the dependence on $a_0$
of the nuclear and kaonic radii %$R_N$, $R_K$
$R_n$, $R_p$, $R_N$, $R_K$
(upper panels)
and the deformation parameter $\beta_2$
%$\beta_2^{(p)}$, $\beta_2^{(n)}$, $\beta_2$
(central panels)
at the minimum of the PES for the three kaonic nuclei.
The shrinking of the nuclear cores
and in particular of the trapped kaon wave function
is clearly evident,
with an associated reduction of $\beta_2$.
The effect is more dramatic for the lightest nuclei,
where the deformation vanishes completely for large enough $a_0$,
as seen in Fig.~\ref{f:def}.
In the insets
%for $^{\,8}_K$Be
we emphasize the different behavior of $R_n$ and $R_p$
in the case of the asymmetric KNI $x_0=1$.

Finally,
an interesting question related to the deformation phenomenon
is the modification of the kaon removal energy due to this effect, i.e,
the quantity
\be
 \Delta B_K \equiv B_K^\text{def.} - B_K^\text{nondef.}
%\\\nonumber
 = \Delta E(^AZ) - \Delta E(^A_KZ)
\ ,\quad
\Delta E= E^\text{def.} - E^\text{nondef.}
\:,
\label{e:db}
\ee
obtained from comparing the results of 2D and 1D calculations.
Fig.~\ref{f:ba} (lower panels)
shows the dependence of this quantity on $a_0$
for the three nuclei.
Without any deformation of the kaonic nucleus,
one would have a constant
$\Delta B_K = \Delta E(^AZ) < 0$
due to the deformation of the core nucleus only,
which is indicated by horizontal lines in the figure.
The fact that the kaonic nucleus is deformed increases the removal
energy relative to this value,
but the total result remains negative.
The overall result is small,
of the order of the depths of the deformation minima.

%-------------------------------------------------------------------------------
\section{Conclusions}

We studied the density dependence of the $K^-$ nuclear potential and
properties of kaonic nuclei using a 2D SHF model with a simple $KN$ Skyrme force,
which self-consistently accounts for
the modification of the nuclear core due to the inserted kaon.
We confirmed the existence of instabilities related to the
unrestrained increase of central nucleon and kaon densities,
which currently limits the maximum kaon binding in this approach
to not much more than 150 MeV
for heavy nuclei with the symmetric KNI ($x_0=0$)
and much less
for light nuclei and/or the asymmetric KNI ($x_0=1$).
This demonstrates that it is premature to discuss the structure of
very strongly bound kaonic nuclei,
before the properties of the $NN$ and $KN$ interactions at high density
are reliable known and theoretically under control.

We also studied the shape of $K^-$ nuclei with a deformed nuclear core
in the ground state
and demonstrated a shrinking of the overall size of the nucleus,
together with a slight reduction of the quadrupole deformation
that might even vanish completely for light nuclei and a very strong KNI.

In the future,
completing the construction of the $KN$ Skyrme force
and including the imaginary part of the kaon optical potential,
will render the approach more reliable and predictive.
Hopefully also accurate experimental data on the kaon binding
will become available and allow to fit the parameters of the KNI
for this purpose.

%-------------------------------------------------------------------------------
\section*{Acknowledgments}

We thank Ji-Wei Cui for suggestive discussions.
This work was supported by the National Science Foundation of China
under contract Nos.~11775081 and 11875134,
and the Natural Science Foundation of Shanghai under contract No.~17ZR1408900.

%\vskip-5mm
%-------------------------------------------------------------------------------

\end{document}